\documentclass[aps,pre,amssymb, nofootinbib]{revtex4}\usepackage{latexsym}
\begin{document}
\preprint{MIT-CTP 3810}

\title{Papapetrou energy-momentum tensor for Chern-Simons modified gravity}
\author{
David Guarrera}
\email{guarrera@mit.edu}
\affiliation{Center for Theoretical Physics,
Massachusetts Institute of Technology, Cambridge, MA 02139}
\author{A. J. Hariton}
\email{hariton@lns.mit.edu}
\affiliation{Center for Theoretical Physics,
Massachusetts Institute of Technology, Cambridge, MA 02139}
\date{\today}

\begin{abstract} We construct a conserved, symmetric energy-momentum (pseudo-)tensor for Chern-Simons modified gravity, thus demonstrating that the theory is Lorentz invariant. The tensor is discussed in relation to other gravitational energy-momentum tensors and analyzed for the Schwarzschild, Reissner-Nordstrom, and Friedmann-Robertson-Walker solutions. To our knowledge this is the first confirmation that the Reissner-Nordstrom and Friedmann-Robertson-Walker metrics are solutions of the modified theory. \\ \\ \small MIT-CTP 3810\end{abstract}

\maketitle

\section{Introduction}
The possibility of modifying a four dimensional theory with a three dimensional Chern-Simons (CS) term was first investigated in \cite{Carroll:1989vb}, where such a term was added to electrodynamics. There, it was found that the extra term created a birefringence of the vacuum, leading to plane waves traveling with two polarizations whose velocities differ from $c$ (Lorentz violation) and from each other (parity violation).

In ensuing work \cite{Jackiw:2003pm}, a similar modification of General Relativity (GR) was proposed. In order to carry out such a construction for gravity, one must decide how to embed a three dimensional CS term into four dimensional GR. This is done with the aid of an embedding coordinate, $v_\mu$. In contrast to CS electrodynamics, there is no birefringence of the vacuum, though there are parity violating effects that cause gravitational wave polarizations to carry different intensities. Moreover, it was argued that the theory allows the construction of a symmetric and conserved two-index object which could serve as an energy-momentum (pseudo-)tensor. \footnote{The two indexed objects of this paper do not correctly transform as tensors and for this reason are referred to as pseudotensors. See the discussion in Section \ref{secion} for more on this issue.  Henceforth,  all references to gravitational energy-momentum tensors should be understood to be references to pseudotensors.} For these reasons it was suggested that the apparent Lorentz violation of the theory is ``dynamically suppressed." 

In this paper we use the Noether/Belinfante procedure to construct a symmetric, conventionally conserved energy-momentum tensor for CS modified gravity. The existence of such a tensor signals the absence of Lorentz violation in the theory. The methods are similar to those used in the construction of the so-called Papapetrou energy-momentum tensor for GR in \cite{Papapetrou:1948jw} and \cite{Bak:1993us}. We find that while the constructed tensor initially appears not to be conserved, a subsidiary condition on solutions of the theory forces the tensor's non-vanishing divergence to zero.

\section{A Brief Review of CS Modified Gravity}

This section is a brief review of \cite{Jackiw:2003pm}, where four dimensional CS modified gravity was examined. The Lagrangian density of the theory is

\begin{equation} \mathcal{L}=\frac{1}{16 \pi G}(\sqrt{-g} R + \frac{1}{4} \theta(x) {}^{\ast}R R), \label{laone}  \end{equation}
where $\theta(x)$ is a prescribed, non-dynamical external field that breaks diffeomorphism symmetry,  ${}^{\ast}R R \equiv {}^{\ast}R^{\sigma \phantom \tau \mu \nu}_{ \phantom \sigma \tau} R^{\tau}_{\phantom \tau \sigma \mu \nu}$, and ${}^{\ast}R^{\sigma \phantom \tau \mu \nu}_{\phantom \sigma \tau} \equiv \frac{1}{2} \epsilon^{\mu \nu \alpha \beta} R^{\sigma \phantom \tau}_{\phantom \sigma  \tau \alpha \beta}$. One generally takes $\theta(x)=v_\sigma x^\sigma$, and timelike $v_\mu=(\frac{1}{\mu}, 0, 0, 0)$, with $\frac{1}{\mu}$ constant. This ensures the persistence of some familiar GR solutions and also maintains the close analogy with 3 dimensional CS theories. 
We note that ${}^{\ast}R R=2 \partial_\mu K^\mu$ is a total derivative, where
\begin{equation} K^\mu=2 \epsilon^{\mu \alpha \beta \gamma} [\frac{1}{2} \Gamma^\sigma_{\alpha \tau} \partial_\beta \Gamma^\tau_{\gamma \sigma} + \frac{1}{3} \Gamma^\sigma_{\alpha \tau} \Gamma^\tau_{\beta \eta} \Gamma^\eta_{\gamma \sigma}],  \end{equation} 
and $\Gamma_{\alpha \beta}^\gamma$ is the Christoffel connection. Upon integrating the Lagrangian by parts, it may be rewritten 
\begin{equation} \mathcal{L}'=\mathcal{L}_{EH}  - \frac{1}{32 \pi G} (v_{\sigma} K^\sigma),  \label{latwo}\end{equation}
where $\mathcal{L}_{EH}$ is the Einstein-Hilbert Lagrangian. Thus the translation non-invariance of (\ref{laone}) is confined to a surface term in the action.
By varying the Lagrangian (plus matter degrees of freedom) with respect to $g$, one finds the equations of motion

\begin{equation}G^{\mu \nu} + C^{\mu \nu}=8 \pi G T^{\mu \nu}.  \label{eom}  \end{equation}
$G^{\mu \nu}$ is the usual Einstein tensor, $T^{\mu \nu}$ is the energy-momentum tensor for matter and $C^{\mu \nu}$ is the following four dimensional analogue of the Cotton tensor,  

\begin{equation} C^{\mu \nu}=-\frac{1}{2 \sqrt{-g}}[v_\sigma (\epsilon^{\sigma \mu \alpha \beta} \nabla_\alpha R^\nu_\beta + \epsilon^{\sigma \nu \alpha \beta} \nabla_\alpha R^\mu _\beta) - v_\alpha \Gamma^\alpha_{\sigma \tau}({}^{\ast}R^{\tau \mu \sigma \nu}+ {}^{\ast} R^{\tau \nu \sigma \mu})].  \end{equation}
Taking the divergence of this equation gives

$$\nabla_{\mu} C^{\mu \nu}=\frac{1}{8 \sqrt{-g} } v^{\nu} {}^{\ast}R R.  $$
However, via the Bianchi identity, $\nabla_\mu G^{\mu \nu}=0$ and for diffeomorphism-invariant matter terms, $\nabla_\mu T^{\mu \nu}=0$. Therefore we have a consistency condition for solutions to (\ref{eom}):

 \begin{equation} {}^{\ast}R R=0 . \label{const} \end{equation}

CS modified gravity theories have been studied as models for parity violation \cite{Lue:1998mq} and leptogenesis \cite{Alexander:2004us}, \cite{Lyth:2005jf} in the early universe. CS models have also been used as effective theories where the CS term is radiatively generated via fermions coupling to gravity in a parity violating way \cite{Mariz:2004cv}. 

\subsection{Solutions}
The Schwarzschild, Reissner-Nordstrom, and all Friedmann-Robertson-Walker (FRW) metrics have vanishing $C_{\mu \nu}$ in their usual coordinatizations and hence are solutions of CS modified gravity. However, the most general black hole solution, the Kerr metric, has non-vanishing $C_{\mu \nu}$ and is not a solution of CS modified gravity. This can be seen easily by noting ${}^{\ast}R R\neq0$ for the Kerr metric. The discovery of an appropriate generalization of the Kerr metric is an outstanding problem. The only non-GR ($C_{\mu \nu} \neq 0$) solutions yet discovered are gravitational waves  \cite{Jackiw:2003pm}. Unlike their GR counterparts, parity violating effects cause the two CS modified wave polarizations to travel with different intensities.   

Though CS modified gravity is not invariant under general diffeomorphisms, we may identify a smaller equivalence class of coordinate transformations. In \cite{Jackiw:2003pm} it is shown that constant shifts in time and arbitrary space reparametrizations are symmetries of the CS modified action. Thus we may view solutions related by these coordinate transformations as identical. 

\section{A Word on Gravitational energy-momentum Tensors}  \label{secion}

The issue of ordinarily conserved energy-momentum tensors for gravity has been controversial since the birth of GR. Einstein's own ``tensor" was non-symmetric and not a tensor (almost all, including the type derived in this paper are coordinate dependent ``pseudotensors"), drawing criticism from leading physicists of the day (these criticisms are nicely reviewed in \cite{Chandra}). The problem with a local definition of gravitational energy-momentum is that there always exists a coordinate system where the energy and momentum densities vanish at a point, viz. Riemannian normal coordinates. In GR, local energy momentum can be ``gauged" away.  Since Einstein's pseudotensor, various other pseudotensors have appeared in the literature including those of Tolman  \cite{Tolman}, Landau and Lifshitz \cite{doublel}, Papapetrou \cite{Papapetrou:1948jw}, \cite{Bak:1993us}, Weinberg \cite{Weinberg} and M{\o}ller \cite{Moller}. None but M{\o}ller's are coordinate invariant. Also, many involve an auxiliary Minkowski metric $\eta=\mbox{diag}(-1,1,1,1)$, and all but M{\o}ller's give physically sensible results only when restricted to ``quasi-Cartesian" coordinate systems. (``Quasi-Cartesian" is defined as $d s^2 \rightarrow -dt^2+dx^2+dy^2 +dz^2$ asymptotically or, less restrictively, that all four coordinates be non-compact. This definition is still a point of debate and is of fundamental importance when one tries to apply these pseudotensors to cosmological models.)

There are other problems. Aguirregabiria, et al. \cite{Aguirregabiria:1995qz} have shown that the Einstein, Tolman, Landau and Lifshitz, Papapetrou and Weinberg (ETLLPW) pseudotensors are identical for any Kerr-Schild metric. Many standard solutions can be put in Kerr-Schild form, including the Schwarzschild, Reissner-Nordstrom, Kerr, and Kerr-Newman metrics. However, Virbhadra later showed \cite{Virbhadra:1998kd} that ETLLPW each give different results for the energy contained in a sphere of radius $r$ when applied to the most general non-static, spherically symmetric metric in ``Schwarzschild Cartesian coordinates" ($(r, \theta, \phi) \rightarrow (x,y,z)$ in the usual way). Furthermore, the Einstein pseudotensor is the only one whose result for the energy contained in a sphere of radius $r$ agrees for the Schwarzschild metric when compared in Kerr-Schild coordinates and Schwarzschild Cartesian coordinates.  

For a short time, it seemed that these problems might be solved by using the concept of quasi-local energy momentum: energy and momentum associated to closed, spacelike 2-surfaces surrounding a region \cite{Penrose:1982wp}. In this way, some of the issues that plague local, pointwise definitions of gravitational energy-momentum are circumvented. However, Bergqvist has investigated seven different definitions of quasi-local mass \cite{Bergqvist}. Computing them on cross sections of the event horizon in a Kerr spacetime and spheres in a Reissner-Nordstrom spacetime, he found that no two of the seven definitions give the same result. 

Despite these problems, though, many authors have given compelling physical arguments for the existence of truly localizable gravitational energy-momentum \cite{Coop}, \cite{Bondi}. These details remain largely unresolved in GR and all other metric theories of gravity. They have been famously confusing for a long time. N. Rosen calculated the Einstein and Landau-Lifshitz pseudotensors for cylindrical gravitational waves \cite{Rosen}. He erroneously used cylindrical coordinates and found that the waves carry zero energy and momentum. These results had many, including Einstein, briefly convinced that gravitational waves did not exist and were merely a coordinate artifact. 

At the very least, it is widely agreed that while these issues are unclear locally, all pseudotensors generally give correct results when applied at infinity for asymptotically flat spacetimes in quasi-Cartesian coordinates (there are important exceptions, see e.g. \cite{ADM1}). 

We shall restrict the calculations of energy and momentum to infinity. The existence of the local tensor signals Lorentz invariance of the theory, but the tensor itself will never be used for any physical computation. Though we are skeptical, there might be some local sense in which our energy-momentum tensor is valid, perhaps when restricted to Kerr-Schild metrics, for example. 

The energy-momentum tensor that we derive for CS modified gravity is closely analogous to the Papapetrou tensor of GR. Like the Papapetrou tensor, it is nicely derived by a Noether argument followed by a Belinfante symmetrization. 

\section{The Belinfante Procedure for Lorentz Invariant Theories}

 We assume a Lorentz invariant Lagrangian of some field (possibly non-scalar, Lorentz indices are suppressed) $\phi$, and show how to construct a symmetric, conserved energy-momentum tensor. We consider the possibility that the Lagrangian involves second derivatives, $\mathcal{L}=\mathcal{L}(\phi, \partial \phi, \partial \partial \phi)$. In terms of the quantities,  $\pi \equiv \frac{\partial \mathcal{L}}{\partial \phi}$, $\pi^\mu \equiv \frac{\partial \mathcal{L} }{\partial (\partial_\mu \phi)}$, and $\pi^{\mu \nu} \equiv \frac{\partial \mathcal{L} }{\partial (\partial_{\mu} \partial{\nu} \phi)}$, the (non-symmetric) canonical tensor derived via Noether's theorem is

\begin{equation} \theta^{\mu \alpha}_C= \pi^{\mu} \partial^{\alpha} \phi + \pi^{\mu \nu} \partial_{\nu}\partial^{\alpha} \phi - \partial_{\nu} \pi^{\mu \nu} \partial^{\alpha} \phi - \eta^{\mu \alpha} \mathcal{L} \label{canonical} \end{equation}
and the equations of motion are

\begin{equation} \partial_{\mu} \pi^{\mu}= \pi + \partial_{\mu} \partial_{\nu} \pi^{\mu \nu} . \end{equation}
One seeks to decompose the tensor as\footnote{We use the conventions $T^{[ab]} \equiv \frac{1}{2}(T^{ab}-T^{ba})$ and $T^{(ab)} \equiv \frac{1}{2}(T^{ab}+T^{ba})$.} 

\begin{equation} \theta^{\mu \alpha}_C=\theta^{\mu \alpha}_B + \partial_{\nu} X^{[\nu \mu] \alpha},  \label{teqn} \end{equation}
with $\theta^{\mu \alpha}_B$ symmetric and $\partial_{\nu} X^{[\nu \mu] \alpha}$ a manifestly conserved, so-called  ``superpotential." Then  $\theta^{\mu \alpha}_B$ will be our conserved, symmetric Belinfante improved energy-momentum tensor. It is well known that this can generally be done in Lorentz invariant theories. The result is 

\begin{eqnarray}  \theta^{\mu \alpha}_B & =&  \pi^{(\mu} \partial^{\alpha)} \phi -  2 \mbox{ }  \partial_\nu \pi^{\nu (\mu} \partial^{\alpha)} \phi- \eta^{\mu \alpha} \mathcal{L}+ \pi^{(\mu} \Sigma^{\alpha )\nu } \phi \nonumber \\
{}&{}&   + \partial_\nu(\pi^{\alpha \mu} \partial^\nu \phi) - \partial_\nu( \partial_\sigma \pi^{\sigma(\mu } \Sigma^{ \alpha) \nu} \phi- \pi^{\sigma(\mu} \Sigma^{ \alpha) \nu} \partial_\sigma \phi) .   \label{btensor} \end{eqnarray}
$\Sigma^{\alpha \beta}$ are the spin matrices for $\phi$ with Lorentz indices suppressed.  By (\ref{teqn}) 

\begin{equation} \partial_\mu \theta^{\mu \alpha}_B=\partial_\mu \theta^{\mu \alpha}_C=0 \end{equation}
i.e., the symmetric tensor is conserved. 

It should be noted that the existence of a conserved, symmetric energy-momentum tensor implies Lorentz invariance of the S-matrix (\cite{WeinbergQFT} sec. 7.4). To see this, we note that the tensor density

\begin{equation}
{\mathcal M}^{\lambda\mu\nu}=x^{\mu}\theta^{\lambda\nu}_B-x^{\nu}\theta^{\lambda\mu}_B,
\label{tensorM}\end{equation}
is conserved in the sense that $\partial_{\lambda}{\mathcal M}^{\lambda\mu\nu}=0$. Thus, we obtain the time-independent tensor

\begin{equation}
J^{\mu\nu}=\int {\mathcal M}^{0\mu\nu}d^3x=\int d^3x(x^{\mu}\theta^{0\nu}_B-x^{\nu}\theta^{0\mu}_B),
\label{tensorJ}\end{equation}
in addition to the time-independent energy-momentum coordinates

\begin{equation}
P^{\mu}=\int \theta^{0\mu}_Bd^3x,
\label{tensorP}\end{equation}
where $H=P^0$ is the energy. If we define the ``rotation'' generators as $J_k=\frac{1}{2}\varepsilon_{ijk}J^{ij}$ and the ``boost'' generators as $K_k=J^{k0}$, we obtain the following commutation relations
\begin{equation}
[H,J_k]=0,\quad [P_j,J_i]=-i\varepsilon_{ijk}P^k,\quad [H,K_k]=-iP_k,\quad [P^j,K_k]=-i\delta^{j}_{k}H,
\label{commut}\end{equation}
which imply Lorentz invariance of the S-matrix (\cite{WeinbergQFT} sec. 3.3).

\section{Energy-momentum Tensor for CS Modified Gravity}

In CS modified gravity, we have the Lagrangian $\mathcal{L}= \mathcal{L}_{EH} + \Delta \mathcal{L}$, where $\mathcal{L}_{EH}$ is the usual, Lorentz-invariant,  Einstein-Hilbert term and  $\Delta \mathcal{L}= \frac{1}{4} (v_{\sigma} x^{\sigma}) {}^{\ast}R R$. We use the abbreviated notation $\mathcal{L}=\mathcal{L}(\phi, \partial \phi, \partial \partial \phi)$, where $\phi$ is understood to be the spacetime metric with indices suppressed. Though we no longer have manifest translational invariance, we can still construct a conserved (non-symmetric) energy-momentum tensor because the translation non-invariant part of the Lagrangian leads to a surface term. Under some infinitesimal transformation

\begin{equation} \delta \mathcal{L}= \pi \delta \phi + \pi^\mu \partial_\mu(\delta \phi)+ \pi^{\mu \nu} \partial_\mu \partial_\nu (\delta \phi). \label{tensorone} \end{equation}
Using the equations of motion, it follows that

\begin{equation} \delta \mathcal{L}=\partial_\mu [\pi^\mu \delta \phi + \pi^{\mu \nu} \partial_\nu \delta \phi-\partial_\nu \pi^{\mu \nu} \delta \phi].  \label{tensortwo} \end{equation}
For an infinitesimal translation, $\delta \phi=\partial^\alpha \phi$, and equation (\ref{tensorone}) gives

\begin{eqnarray} \delta \mathcal{L} &=& \pi \partial^\alpha \phi + \pi^\mu \partial_\mu \partial^\alpha \phi + \pi^{\mu \nu} \partial_\mu \partial_\nu \partial^\alpha \phi \nonumber \\
{}&=& \partial^\alpha \mathcal{L} - \frac{\mbox{d} \mathcal{L}}{\mbox{d} x_\alpha} \nonumber \\
{}&=&\partial^\alpha \mathcal{L} -\frac{v^\alpha}{4}  {}^{\ast}R R \nonumber \\
{}&=&\partial_\mu[\eta^{\alpha \mu} \mathcal{L} -\frac{v^\alpha}{2} K^\mu]. \label{tensorthree}\end{eqnarray}
Equating (\ref{tensortwo}) and (\ref{tensorthree}), 

\begin{equation} \partial_\mu[\pi^\mu \partial^\alpha \phi+ \pi^{\mu \nu} \partial_\nu \partial^\alpha \phi - \partial_\nu \pi^{\mu \nu} \partial^\alpha \phi - \eta^{\alpha \mu} \mathcal{L} + \frac{v^\alpha}{2} K^\mu]=0, \end{equation}
and so we have a conserved energy-momentum tensor
\begin{equation} \theta^{\mu \alpha}=\pi^\mu \partial^\alpha \phi+ \pi^{\mu \nu} \partial_\nu \partial^\alpha \phi - \partial_\nu \pi^{\mu \nu} \partial^\alpha \phi - \eta^{\alpha \mu} \mathcal{L} + \frac{v^\alpha}{2} K^\mu. \end{equation}
We label this as $\theta^{\mu \alpha}=\theta^{\mu \alpha}_C + \frac{v^\alpha}{2} K^\mu$, where $\theta^{\mu \alpha}_C$ is the usual formula (\ref{canonical}) for the energy-momentum tensor for translationally invariant Lagrangians.  It should be noted that the $\pi$'s present in this equation are derivatives of the full, Lorentz non-invariant Lagrangian. By implementing a similar Belinfante procedure as in the previous section, one can massage $\theta^{\mu \alpha}_C$ such that 

\begin{equation} \theta^{\mu \alpha}_C=\theta^{\mu \alpha}_B + \partial_\nu X^{[\nu \mu] \alpha} + A^{\mu \alpha}, \end{equation}
where $\theta^{\mu \alpha}_B$ is as in equation (\ref{btensor}), $\partial_\nu X^{[\nu \mu] \alpha} $ is a superpotential, and $A^{\mu \alpha}$ contains only terms that are proportional to ${}^{\ast}R R$ and derivatives of ${}^{\ast}R R$. Because of the dynamical consistency condition (\ref{const}), $A^{\mu \alpha}=0$ for solutions and 
\begin{equation} \theta^{\mu \alpha}=\theta^{\mu \alpha}_B +  \partial_\nu X^{[\nu \mu] \alpha} +\frac{v^\alpha}{2} K^\mu \end{equation}
and

\begin{equation} 0=\partial_\mu \theta^{\mu \alpha}=\partial_\mu \theta^{\mu \alpha}_B + \frac{v^\alpha}{2} \partial_\mu K^\mu.  \end{equation}
By (\ref{const}) $\partial_\mu K^\mu=0$ , and so $\theta^{\mu \alpha}_B$ is a conserved, symmetric, energy-momentum tensor. We again note that this Papapetrou pseudotensor, like many other gravitational pseudotensors, necessitates a background Minkowski metric $\eta=(-1,1,1,1)$, and therefore should only be used in quasi-Cartesian, asymptotically flat coordinates. The tensor is given by replacing $\phi$ in equation (\ref{btensor}) by the spacetime metric, $g_{ab}$:

\begin{eqnarray}  \theta^{\mu \alpha}_B & =&  \pi^{\{a b\} (\mu} \partial^{\alpha)} g_{a b} - 2 \mbox{ }\partial_\nu \pi^{\{a b\} \nu (\mu} \partial^{\alpha)} g_{a b}- \eta^{\mu \alpha} \mathcal{L} +  (\partial_\nu \pi^{\{a b\}(\mu }_{\phantom b } \Sigma^{\alpha) \nu }_{a b} g)   \nonumber \\
{}&{}& +  \partial_\nu(\pi^{\{a b\} \alpha \mu} \partial^\nu g_{a b}) - \partial_\nu( \partial_\sigma \pi^{\{a b\} \sigma  (\mu }_{\phantom b} \Sigma^{\alpha) \nu }_{a b} g- \pi^{\{a b\} \sigma (\mu  }_{\phantom b} \Sigma^{\alpha) \nu }_{a b} \partial_\sigma g), \label{letensor}  \end{eqnarray}
where   $\pi^{\{a b\} \mu }= \frac{\partial \mathcal{L}}{\partial (\partial_\mu g_{ab})}$ and similarly for $\pi^{\{a b\} \mu \nu }$. The gravitational spin matrices are

\begin{equation} \Sigma^{\nu \mu}_{a b} g=(\eta^{\mu \sigma} \delta^{\nu}_a -\eta^{\nu\sigma} \delta^{\mu}_a) g_{\sigma b} +( \eta^{\mu \sigma} \delta^{\nu}_b -\eta^{\nu\sigma} \delta^{\mu}_b) g_{\sigma a}. \end{equation}
By linearity $\pi^{\{ab\} \mu} =\pi^{\{ab\} \mu}_{EH} + \pi^{\{ab\} \mu}_{CS}$, where

$$\pi^{\{a b\} \nu}_{EH} \equiv \frac{\partial (\frac{1}{16 \pi G} \sqrt{-g} R)}{\partial (\partial_\nu g_{ab})}\mbox{ and } \pi^{\{a b\} \nu}_{CS} \equiv \frac{\partial (\frac{1}{64 \pi G} \theta(x) {}^{\ast}R R)}{\partial (\partial_\nu g_{ab})}, $$
and similarly for $ \pi^{\{ab\} \rho \nu }$. It is straightforward (though lengthy) to calculate that

\begin{equation} \pi^{\{a b\} \nu}_{CS} =\frac{\theta(x)}{32 \pi G}(-\Gamma^\nu_{\tau \mu} {}^*R^{b \tau \mu a} -\Gamma^\nu_{\tau \mu} {}^*R^{a \tau \mu b} + \Gamma^b_{\tau \mu} {}^*R^{\nu \tau \mu a} + \Gamma^a_{\tau \mu} {}^*R^{\nu \tau \mu b} +\Gamma^a_{\tau \mu} {}^*R^{b \tau \mu \nu} + \Gamma^b_{\tau \mu} {}^*R^{a \tau \mu b}) \end{equation}
and 

\begin{equation} \pi^{\{ab\} \rho \nu }_{CS}=\frac{\theta(x)}{64 \pi G} ({}^*R^{b \rho \nu a}+ {}^*R^{a \rho \nu b} + {}^*R^{b \nu \rho a} + {}^*R^{a \nu \rho b}). \end{equation}

We would like to reiterate that although we have formulated a symmetric, conserved local energy-momentum pseudotensor, we do not believe in any local physical intepretation. The utility of this tensor is twofold: it can be used to calculate the total energy, momentum and angular momentum for spacetimes of the theory and its existence signals that a seemingly Lorentz violating theory is actually Lorentz invariant.

\section{Comparison With the Weinberg Tensor}

We briefly digress on another method for computing an energy-momentum tensor in CS modified gravity that was investigated in  \cite{Jackiw:2003pm}.   The vacuum equations of motion are

\begin{equation}G_{\mu \nu} + C_{\mu \nu}=0 . \end{equation}
Now, take a quasi-Cartesian coordinate system with $h_{\mu \nu}\equiv g_{\mu \nu} - \eta_{\mu \nu}$. Expanding the above equation in powers of $h$, 

\begin{equation} G_{\mu \nu}^{(1)} + C_{\mu \nu}^{(1)} =8 \pi G t_{\mu \nu}, \label{wein2} \end{equation}
where
\begin{equation} t_{\mu \nu} \equiv -\frac{1}{8 \pi G}[G_{\mu \nu} + C_{\mu \nu}-G_{\mu \nu}^{(1)} - C_{\mu \nu}^{(1)}] \label{wein} \end{equation} 
and the superscripts denote the order in $h$. The tensor $t_{\mu \nu}$ has most of the properties we might want from a gravitational energy-momentum tensor: it is symmetric, ordinarily conserved (because of the linear Bianchi identity and the linear version of (\ref{const})), and quadratic in $h$ (though it is not, as usual, coordinate invariant). In GR, the energy-momentum tensor derived as (\ref{wein}) is referred to as the Weinberg tensor (see \cite{Weinberg}). To compute the total energy, momentum or angular momentum of a gravitational system, we may integrate the left hand side of (\ref{wein2}), which is also sometimes referred to as the Weinberg tensor. One can explicitly verify that doing so gives the Arnowitt, Deser and Misner (ADM) total energy and momentum \cite{Arnowitt:1962hi}. 

We now demonstrate that our CS modified Papapetrou tensor gives the same result for the total energy, momentum, and angular momentum of a spacetime as (\ref{wein}). Taking an asymptotically flat spacetime with $h=\mbox{O}(\frac{1}{r})$, we can expand our expression for the energy-momentum tensor, equation (\ref{letensor}), to lowest order. The only finite contributions to the total energy, momentum, and angular momentum of a spacetime will be the lowest order terms in $\frac{1}{r}$.  These are $\mbox{O}(\frac{1}{r^3})$ for energy and linear momentum and $\mbox{O}(\frac{1}{r^4})$ for angular momentum. All higher orders will die off at infinity. It is easily derived that 

\begin{equation} \theta^{\mu \alpha}_B(\pi^{\{ab\} \beta}, \pi^{\{ab\} \beta \sigma}, g_{ab})=\theta^{\mu \alpha}_B(\pi^{\{ab\} \beta}_{EH}, \pi^{\{ab\} \beta \sigma}_{EH}, g_{ab})+\mbox{O}(\frac{1}{r^5}) . \end{equation}
$\theta^{\mu \alpha}_B(\pi^{\{ab\} \beta}_{EH}, \pi^{\{ab\} \beta \sigma}_{EH}, g_{ab})$ can be calculated rather straightforwardly, as in \cite{Bak:1993us}, using the equations of motion several times to obtain a nice form. The result is

\begin{equation} \theta^{\mu \alpha}_B(\pi^{\{ab\} \beta}_{EH}, \pi^{\{ab\} \beta \sigma}_{EH}, g_{ab})=-\frac{1}{16 \pi G} \partial_\gamma \partial_\beta \sqrt{-g}[\eta^{\mu \alpha} g^{\gamma \beta} - \eta^{\gamma \alpha} g^{\mu \beta}  + \eta^{\gamma \beta} g^{\mu \alpha} -\eta^{\mu \beta} g^{\gamma \alpha}] + \frac{\sqrt{-g}}{16 \pi G} [\eta^{\alpha \gamma} g^{\mu \beta} C_{\beta \gamma} + \eta^{\mu \gamma} g^{\alpha \beta} C_{\beta \gamma}].  \end{equation} 
We can continue the expansion,

\begin{eqnarray} \theta^{\mu \alpha}_B(\pi^{\{ab\} \beta}, \pi^{\{ab\} \beta \sigma}, g_{ab})&=&\theta^{\mu \alpha}_B(\pi^{\{ab\} \beta}_{EH}, \pi^{\{ab\} \beta \sigma}_{EH}, g_{ab})+\mbox{O}(\frac{1}{r^5}) \nonumber \\ {}&=&-\frac{1}{16 \pi G} \partial_\gamma \partial_\beta \sqrt{-g}[\eta^{\mu \alpha} g^{\gamma \beta} - \eta^{\gamma \alpha} g^{\mu \beta}  + \eta^{\gamma \beta} g^{\mu \alpha} -\eta^{\mu \beta} g^{\gamma \alpha}] \nonumber \\ {}& +& \frac{\sqrt{-g}}{16 \pi G} [\eta^{\alpha \gamma} g^{\mu \beta} C_{\beta \gamma} + \eta^{\mu \gamma} g^{\alpha \beta} C_{\beta \gamma}]+\mbox{O}(\frac{1}{r^5}) \nonumber \\ {}&=&-\frac{1}{16 \pi G} [-\partial^\mu \partial^\alpha h^\lambda_\lambda+\partial_\lambda \partial^\alpha h^{\lambda \mu} + \partial_\lambda \partial^\mu h^{\lambda \alpha} - \Box h^{\mu \alpha} + \eta^{\mu \alpha} \Box h^\lambda_\lambda - \eta^{\mu \alpha} \partial_\lambda \partial_\sigma h^{\lambda \sigma}] \nonumber \\ {}&+& \frac{1}{8 \pi G} C^{\mu \alpha (1)}+\mbox{O}(\frac{1}{r^5}) \nonumber \\ {}&=&\frac{1}{8 \pi G} [G^{\mu \alpha (1)} +C^{\mu \alpha (1)}]+\mbox{O}(\frac{1}{r^5}) , \end{eqnarray}
and so
\begin{equation} \int_V \mbox{d}^3 x \mbox{ } \theta^{\mu \alpha}_B =\frac{1}{8 \pi G}\int_V  \mbox{d}^3 x \mbox{ }  [G^{\mu \alpha (1)} +C^{\mu \alpha (1)}]. \end{equation}
It is also immediate from (\ref{wein2}) and (\ref{wein}) that

\begin{equation} \int_V t^{\mu \alpha}=\frac{1}{8 \pi G}\int_V  \mbox{d}^3 x \mbox{ }  [G^{\mu \alpha (1)} +C^{\mu \alpha (1)}].  \end{equation}
so that the total energy, momentum and angular momentum  of a spacetime is the same calculated with the Papapetrou tensor (e.g. $E=\int_V \mbox{d}^3 x \mbox{ } \theta^{0 0}_B$) as with the Weinberg tensor ($E=\int_V \mbox{d}^3 x \mbox{ } t^{0 0}$). 

\section{Energy-Momentum of Solutions}
The energy-momentum tensor (\ref{letensor}) was calculated for the Schwarzschild and Reissner-Nordstrom solutions using quasi-Cartesian coordinates ($(r, \theta, \phi) \rightarrow (x,y,z)$ in the usual way). We have shown (in this case, with Maple v7) that for each of these solutions all terms in (\ref{letensor}) that involve $\pi_{CS}^{\{ab\}\mu}$ and $\pi_{CS}^{\{ab\} \mu \nu}$ are zero. The energy-momentum tensor evaluated for these solutions is thus unchanged from GR. It is comforting to know that in CS modified gravity a black hole with mass $M$, charge $Q$ and angular momentum zero is still an admissible solution. A generalization of this result to non-zero angular momentum black hole solutions remains an open problem. 

The most general ($k=\{0,-1,1\}$) FRW solution has vanishing $\pi^{\{ab\} \nu}_{CS}$ and $\pi^{\{ab\} \rho \nu}_{CS}$ in ``Cartesian coordinates":

\begin{equation} ds^2=dt^2-a(t)^2 [\frac{kr^2}{1-kr^2}(\frac{x}{r} \mbox{ }dx + \frac{y}{r}\mbox{ } dy+\frac{z}{r} \mbox{ }dz)^2+dx^2+dy^2+dz^2 ] ,\end{equation}
with $r^2=x^2+y^2+z^2$. Therefore, the energy-momentum tensor of FRW models in CS modified gravity is also identical to its value in GR. Beginning with \cite{Rosen:1994vj} various authors have used the ETLLPW pseudotensors in such coordinates to analyze the energy content ($\theta^{0 0}+T^{0 0}$) of both open and closed FRW solutions in GR. The merit of these pseudotensors is debatable in this case, since the spacetimes are not asymptotically flat. Nevertheless, such analyses seem to give reasonable physical results that have been suggested by other coordinate-invariant analyses  (see e.g.  \cite{Katz1}, \cite{Katz:1996nr}). In  \cite{Johri:1995gh}, \cite{Garecki:2006qy} and  \cite{Berman:2006zx} it is shown that ETLLPW all give zero total energy for any finite volume of flat FRW models.  If such calculations turn out to have physical merit, their results are directly applicable to CS modified gravity.

\section{Conclusions}
We have constructed a symmetric, conserved energy-momentum tensor for CS modified gravity and evaluated it on some sample spaces. We might now consider Lagrangian (\ref{laone}) with $\theta$ as a Lagrange multiplier instead of a prescribed field. This theory is now explicitly diffeomorphism and Lorentz invariant because $\theta$ now responds to coordinate transformations,  and therefore the theory admits a symmetric, conserved energy-momentum tensor. Varying with respect to $g$ again gives (\ref{eom}) as an equation of motion, while varying with respect to $\theta$ immediately gives the consistency condition (\ref{const}) as an equation of motion. By making a coordinate transformation we may set $\theta(x,t) \propto t$, and we obtain CS modified gravity as a coordinate choice in this new theory. Thus the ``Lorentz violation" of CS modified gravity is just a choice of coordinates in the new theory. Viewed in this light, it is not very surprising that CS modified gravity does indeed admit a conserved energy-momentum tensor that signals the absence of Lorentz violation.

\begin{acknowledgments}
The authors would like to thank R. Jackiw for introducing them to CS modified gravity, providing the inspiration for this project and for many useful related discussions. They would also like to thank S. Deser and C.W. Misner for useful comments on an earlier draft.  This work is supported by the U.S. Department of Energy under cooperative research agreement No. DEFG02-05ER41360. A. J. H. acknowledges support by the FQRNT of Canada. 
\end{acknowledgments}

\bibliography{cspaperbib}

\begin{thebibliography}{29}
\expandafter\ifx\csname natexlab\endcsname\relax\def\natexlab#1{#1}\fi
\expandafter\ifx\csname bibnamefont\endcsname\relax
  \def\bibnamefont#1{#1}\fi
\expandafter\ifx\csname bibfnamefont\endcsname\relax
  \def\bibfnamefont#1{#1}\fi
\expandafter\ifx\csname citenamefont\endcsname\relax
  \def\citenamefont#1{#1}\fi
\expandafter\ifx\csname url\endcsname\relax
  \def\url#1{\texttt{#1}}\fi
\expandafter\ifx\csname urlprefix\endcsname\relax\def\urlprefix{URL }\fi
\providecommand{\bibinfo}[2]{#2}
\providecommand{\eprint}[2][]{\url{#2}}

\bibitem[{\citenamefont{Carroll et~al.}(1990)\citenamefont{Carroll, Field, and
  Jackiw}}]{Carroll:1989vb}
\bibinfo{author}{\bibfnamefont{S.~M.} \bibnamefont{Carroll}},
  \bibinfo{author}{\bibfnamefont{G.~B.} \bibnamefont{Field}}, \bibnamefont{and}
  \bibinfo{author}{\bibfnamefont{R.}~\bibnamefont{Jackiw}},
  \bibinfo{journal}{Phys. Rev.} \textbf{\bibinfo{volume}{D41}},
  \bibinfo{pages}{1231} (\bibinfo{year}{1990}).

\bibitem[{\citenamefont{Jackiw and Pi}(2003)}]{Jackiw:2003pm}
\bibinfo{author}{\bibfnamefont{R.}~\bibnamefont{Jackiw}} \bibnamefont{and}
  \bibinfo{author}{\bibfnamefont{S.~Y.} \bibnamefont{Pi}},
  \bibinfo{journal}{Phys. Rev.} \textbf{\bibinfo{volume}{D68}},
  \bibinfo{pages}{104012} (\bibinfo{year}{2003}), \eprint{gr-qc/0308071}.

\bibitem[{\citenamefont{Papapetrou}(1948)}]{Papapetrou:1948jw}
\bibinfo{author}{\bibfnamefont{A.}~\bibnamefont{Papapetrou}},
  \bibinfo{journal}{Proc. Roy. Irish Acad. (Sect. A)}
  \textbf{\bibinfo{volume}{52A}}, \bibinfo{pages}{11} (\bibinfo{year}{1948}).

\bibitem[{\citenamefont{Bak et~al.}(1994)\citenamefont{Bak, Cangemi, and
  Jackiw}}]{Bak:1993us}
\bibinfo{author}{\bibfnamefont{D.}~\bibnamefont{Bak}},
  \bibinfo{author}{\bibfnamefont{D.}~\bibnamefont{Cangemi}}, \bibnamefont{and}
  \bibinfo{author}{\bibfnamefont{R.}~\bibnamefont{Jackiw}},
  \bibinfo{journal}{Phys. Rev.} \textbf{\bibinfo{volume}{D49}},
  \bibinfo{pages}{5173} (\bibinfo{year}{1994}), \eprint{hep-th/9310025}.

\bibitem[{\citenamefont{Lue et~al.}(1999)\citenamefont{Lue, Wang, and
  Kamionkowski}}]{Lue:1998mq}
\bibinfo{author}{\bibfnamefont{A.}~\bibnamefont{Lue}},
  \bibinfo{author}{\bibfnamefont{L.-M.} \bibnamefont{Wang}}, \bibnamefont{and}
  \bibinfo{author}{\bibfnamefont{M.}~\bibnamefont{Kamionkowski}},
  \bibinfo{journal}{Phys. Rev. Lett.} \textbf{\bibinfo{volume}{83}},
  \bibinfo{pages}{1506} (\bibinfo{year}{1999}), \eprint{astro-ph/9812088}.

\bibitem[{\citenamefont{Alexander et~al.}(2006)\citenamefont{Alexander, Peskin,
  and Sheikh-Jabbari}}]{Alexander:2004us}
\bibinfo{author}{\bibfnamefont{S.~H.~S.} \bibnamefont{Alexander}},
  \bibinfo{author}{\bibfnamefont{M.~E.} \bibnamefont{Peskin}},
  \bibnamefont{and} \bibinfo{author}{\bibfnamefont{M.~M.}
  \bibnamefont{Sheikh-Jabbari}}, \bibinfo{journal}{Phys. Rev. Lett.}
  \textbf{\bibinfo{volume}{96}}, \bibinfo{pages}{081301}
  (\bibinfo{year}{2006}), \eprint{hep-th/0403069}.

\bibitem[{\citenamefont{Lyth et~al.}(2005)\citenamefont{Lyth, Quimbay, and
  Rodriguez}}]{Lyth:2005jf}
\bibinfo{author}{\bibfnamefont{D.~H.} \bibnamefont{Lyth}},
  \bibinfo{author}{\bibfnamefont{C.}~\bibnamefont{Quimbay}}, \bibnamefont{and}
  \bibinfo{author}{\bibfnamefont{Y.}~\bibnamefont{Rodriguez}},
  \bibinfo{journal}{JHEP} \textbf{\bibinfo{volume}{0503}}, \bibinfo{pages}{016}
  (\bibinfo{year}{2005}), \eprint{hep-th/0501153}.

\bibitem[{\citenamefont{Mariz et~al.}(2004)\citenamefont{Mariz, Nascimento,
  Passos, and Ribeiro}}]{Mariz:2004cv}
\bibinfo{author}{\bibfnamefont{T.}~\bibnamefont{Mariz}},
  \bibinfo{author}{\bibfnamefont{J.~R.} \bibnamefont{Nascimento}},
  \bibinfo{author}{\bibfnamefont{E.}~\bibnamefont{Passos}}, \bibnamefont{and}
  \bibinfo{author}{\bibfnamefont{R.~F.} \bibnamefont{Ribeiro}},
  \bibinfo{journal}{Phys. Rev.} \textbf{\bibinfo{volume}{D70}},
  \bibinfo{pages}{024014} (\bibinfo{year}{2004}), \eprint{hep-th/0403205}.

\bibitem[{\citenamefont{Chandrasekhar and Ferrari}(1991)}]{Chandra}
\bibinfo{author}{\bibfnamefont{S.}~\bibnamefont{Chandrasekhar}}
  \bibnamefont{and} \bibinfo{author}{\bibfnamefont{V.}~\bibnamefont{Ferrari}},
  \bibinfo{journal}{Proc. R. Soc. London} \textbf{\bibinfo{volume}{A435}},
  \bibinfo{pages}{645} (\bibinfo{year}{1991}).

\bibitem[{\citenamefont{Tolman}(1934)}]{Tolman}
\bibinfo{author}{\bibfnamefont{R.}~\bibnamefont{Tolman}},
  \bibinfo{journal}{Relativity, Thermodynamics and Cosmology, Oxford Univ.
  Press, London}  (\bibinfo{year}{1934}).

\bibitem[{\citenamefont{Landau and Lifshitz}(1987)}]{doublel}
\bibinfo{author}{\bibfnamefont{L.}~\bibnamefont{Landau}} \bibnamefont{and}
  \bibinfo{author}{\bibfnamefont{E.}~\bibnamefont{Lifshitz}},
  \bibinfo{journal}{The Classical Theory of Fields, Pergamon Press, London}
  (\bibinfo{year}{1987}).

\bibitem[{\citenamefont{Weinberg}(1972)}]{Weinberg}
\bibinfo{author}{\bibfnamefont{S.}~\bibnamefont{Weinberg}},
  \bibinfo{journal}{Gravitation and Cosmology: Principles and Applications of
  the General Theory of Relativity, John Wiley and Sons, Inc., New York}
  (\bibinfo{year}{1972}).

\bibitem[{\citenamefont{M{\o}ller}(1958)}]{Moller}
\bibinfo{author}{\bibfnamefont{C.}~\bibnamefont{M{\o}ller}},
  \bibinfo{journal}{Ann. Phys.} \textbf{\bibinfo{volume}{4}},
  \bibinfo{pages}{347} (\bibinfo{year}{1958}).

\bibitem[{\citenamefont{Aguirregabiria
  et~al.}(1996)\citenamefont{Aguirregabiria, Chamorro, and
  Virbhadra}}]{Aguirregabiria:1995qz}
\bibinfo{author}{\bibfnamefont{J.~M.} \bibnamefont{Aguirregabiria}},
  \bibinfo{author}{\bibfnamefont{A.}~\bibnamefont{Chamorro}}, \bibnamefont{and}
  \bibinfo{author}{\bibfnamefont{K.~S.} \bibnamefont{Virbhadra}},
  \bibinfo{journal}{Gen. Rel. Grav.} \textbf{\bibinfo{volume}{28}},
  \bibinfo{pages}{1393} (\bibinfo{year}{1996}), \eprint{gr-qc/9501002}.

\bibitem[{\citenamefont{Virbhadra}(1999)}]{Virbhadra:1998kd}
\bibinfo{author}{\bibfnamefont{K.~S.} \bibnamefont{Virbhadra}},
  \bibinfo{journal}{Phys. Rev.} \textbf{\bibinfo{volume}{D60}},
  \bibinfo{pages}{104041} (\bibinfo{year}{1999}), \eprint{gr-qc/9809077}.

\bibitem[{\citenamefont{Penrose}(1982)}]{Penrose:1982wp}
\bibinfo{author}{\bibfnamefont{R.}~\bibnamefont{Penrose}},
  \bibinfo{journal}{Proc. Roy. Soc. Lond.} \textbf{\bibinfo{volume}{A381}},
  \bibinfo{pages}{53} (\bibinfo{year}{1982}).

\bibitem[{\citenamefont{Bergqvist}(1992)}]{Bergqvist}
\bibinfo{author}{\bibfnamefont{G.}~\bibnamefont{Bergqvist}},
  \bibinfo{journal}{Class.Quantum Grav.} \textbf{\bibinfo{volume}{9}},
  \bibinfo{pages}{1753} (\bibinfo{year}{1992}).

\bibitem[{\citenamefont{Cooperstock and Sarracino}(1978)}]{Coop}
\bibinfo{author}{\bibfnamefont{F.}~\bibnamefont{Cooperstock}} \bibnamefont{and}
  \bibinfo{author}{\bibfnamefont{R.}~\bibnamefont{Sarracino}},
  \bibinfo{journal}{J. Phys. A: Math. Gen.} \textbf{\bibinfo{volume}{11}},
  \bibinfo{pages}{877} (\bibinfo{year}{1978}).

\bibitem[{\citenamefont{Bondi}(1990)}]{Bondi}
\bibinfo{author}{\bibfnamefont{H.}~\bibnamefont{Bondi}},
  \bibinfo{journal}{Proc. R. Soc. London} \textbf{\bibinfo{volume}{A427}},
  \bibinfo{pages}{249} (\bibinfo{year}{1990}).

\bibitem[{\citenamefont{Rosen}(1956)}]{Rosen}
\bibinfo{author}{\bibfnamefont{N.}~\bibnamefont{Rosen}},
  \bibinfo{journal}{Gravitational Waves, appears in Jubilee of Relativity
  Theory, Birkhauser Verlag, Basel}  (\bibinfo{year}{1956}).

\bibitem[{\citenamefont{Arnowitt et~al.}(1961)\citenamefont{Arnowitt, Deser,
  and Misner}}]{ADM1}
\bibinfo{author}{\bibfnamefont{R.}~\bibnamefont{Arnowitt}},
  \bibinfo{author}{\bibfnamefont{S.}~\bibnamefont{Deser}}, \bibnamefont{and}
  \bibinfo{author}{\bibfnamefont{C.}~\bibnamefont{Misner}},
  \bibinfo{journal}{Phys. Rev.} \textbf{\bibinfo{volume}{122}},
  \bibinfo{pages}{997} (\bibinfo{year}{1961}).

\bibitem[{\citenamefont{Weinberg}(2005)}]{WeinbergQFT}
\bibinfo{author}{\bibfnamefont{S.}~\bibnamefont{Weinberg}},
  \bibinfo{journal}{The Quantum Theory of Fields, Vol. I: Foundations,
  Cambridge University Press, Cambridge}  (\bibinfo{year}{2005}).

\bibitem[{\citenamefont{Arnowitt et~al.}(1962)\citenamefont{Arnowitt, Deser,
  and Misner}}]{Arnowitt:1962hi}
\bibinfo{author}{\bibfnamefont{R.}~\bibnamefont{Arnowitt}},
  \bibinfo{author}{\bibfnamefont{S.}~\bibnamefont{Deser}}, \bibnamefont{and}
  \bibinfo{author}{\bibfnamefont{C.~W.} \bibnamefont{Misner}}
  (\bibinfo{year}{1962}), \eprint{gr-qc/0405109}.

\bibitem[{\citenamefont{Rosen}(1994)}]{Rosen:1994vj}
\bibinfo{author}{\bibfnamefont{N.}~\bibnamefont{Rosen}}, \bibinfo{journal}{Gen.
  Rel. Grav.} \textbf{\bibinfo{volume}{26}}, \bibinfo{pages}{319}
  (\bibinfo{year}{1994}).

\bibitem[{\citenamefont{Katz et~al.}(1995)\citenamefont{Katz, Bicak, and
  Lynden-Bell}}]{Katz1}
\bibinfo{author}{\bibfnamefont{J.}~\bibnamefont{Katz}},
  \bibinfo{author}{\bibfnamefont{J.}~\bibnamefont{Bicak}}, \bibnamefont{and}
  \bibinfo{author}{\bibfnamefont{D.}~\bibnamefont{Lynden-Bell}},
  \bibinfo{journal}{Mon. Not. R. Astron. Soc.} \textbf{\bibinfo{volume}{272}},
  \bibinfo{pages}{150} (\bibinfo{year}{1995}).

\bibitem[{\citenamefont{Katz et~al.}(1997)\citenamefont{Katz, Bicak, and
  Lynden-Bell}}]{Katz:1996nr}
\bibinfo{author}{\bibfnamefont{J.}~\bibnamefont{Katz}},
  \bibinfo{author}{\bibfnamefont{J.}~\bibnamefont{Bicak}}, \bibnamefont{and}
  \bibinfo{author}{\bibfnamefont{D.}~\bibnamefont{Lynden-Bell}},
  \bibinfo{journal}{Phys. Rev.} \textbf{\bibinfo{volume}{D55}},
  \bibinfo{pages}{5957} (\bibinfo{year}{1997}), \eprint{gr-qc/0504041}.

\bibitem[{\citenamefont{Johri et~al.}(1995)\citenamefont{Johri, Kalligas,
  Singh, and Everitt}}]{Johri:1995gh}
\bibinfo{author}{\bibfnamefont{V.~B.} \bibnamefont{Johri}},
  \bibinfo{author}{\bibfnamefont{D.}~\bibnamefont{Kalligas}},
  \bibinfo{author}{\bibfnamefont{G.~P.} \bibnamefont{Singh}}, \bibnamefont{and}
  \bibinfo{author}{\bibfnamefont{C.~W.~F.} \bibnamefont{Everitt}},
  \bibinfo{journal}{Gen. Rel. Grav.} \textbf{\bibinfo{volume}{27}},
  \bibinfo{pages}{313} (\bibinfo{year}{1995}).

\bibitem[{\citenamefont{Garecki}(2006)}]{Garecki:2006qy}
\bibinfo{author}{\bibfnamefont{J.}~\bibnamefont{Garecki}}
  (\bibinfo{year}{2006}), \eprint{gr-qc/0611056}.

\bibitem[{\citenamefont{Berman}(2006)}]{Berman:2006zx}
\bibinfo{author}{\bibfnamefont{M.~S.} \bibnamefont{Berman}}
  (\bibinfo{year}{2006}), \eprint{gr-qc/0605063}.

\end{thebibliography}

\end{document}